\documentclass[aps,superscriptaddress,twocolumn,twoside,floatfix,pra,nofootinbib,a4paper]{revtex4}
\usepackage{times}
\usepackage{epsfig}
\usepackage{amsfonts}
\usepackage{amsmath}
\usepackage{amssymb}
\usepackage{color}
\usepackage{multirow}
\usepackage[normalem]{ulem}
\usepackage{latexsym}
\usepackage{amsfonts}
\usepackage{mathrsfs}
\usepackage{natbib}
\usepackage{verbatim}
\usepackage[T1]{fontenc}
\usepackage{amsthm}
\usepackage{tabularx,ragged2e}
\usepackage{array,booktabs}
\newcolumntype{C}{>{$}c<{$}}
\AtBeginDocument{
	\heavyrulewidth=.08em
	\lightrulewidth=.05em
	\cmidrulewidth=.03em
	\belowrulesep=.65ex
	\belowbottomsep=0pt
	\aboverulesep=.4ex
	\abovetopsep=0pt
	\cmidrulesep=\doublerulesep
	\cmidrulekern=.5em
	\defaultaddspace=.5em
}

\usepackage[colorlinks=true,linkcolor=blue,citecolor=magenta,urlcolor=blue]{hyperref}

\DeclareMathOperator{\Tr}{tr}

\newcommand{\ket}[1]{|#1\rangle}

\newcommand{\ketbra}[2]{|#1\rangle\langle#2|}

\newcommand{\ba}{\begin{eqnarray}}
	\newcommand{\be}{\begin{equation}}
		\newcommand{\ee}{\end{equation}}

	\newcommand{\ea}{\end{eqnarray}}
\newcommand{\ban}{\begin{eqnarray*}}
	\newcommand{\ean}{\end{eqnarray*}}

\definecolor{orange}{rgb}{1,0.5,0}

\begin{document}


\title{Experimental characterisation of unsharp qubit observables and sequential measurement incompatibility via quantum random access codes}


\author{Hammad Anwer}
\affiliation{Department of Physics, Stockholm University, S-10691 Stockholm, Sweden}

\author{Sadiq Muhammad}
\affiliation{Department of Physics, Stockholm University, S-10691 Stockholm, Sweden}

\author{Walid~Cherifi}
\affiliation{Department of Physics, Stockholm University, S-10691 Stockholm, Sweden}

\author{Nikolai Miklin}
\affiliation{Institute of Theoretical Physics and Astrophysics,	National Quantum Information Center, Faculty of Mathematics, Physics and Informatics, University of Gdansk, 80-952 Gd\'ansk, Poland}

\author{Armin Tavakoli}
\affiliation{D\'epartement de Physique Appliqu\'ee, Universit\'e de Gen\`eve, CH-1211 Gen\`eve, Switzerland}

\author{Mohamed Bourennane}
\affiliation{Department of Physics, Stockholm University, S-10691 Stockholm, Sweden}

\begin{abstract}	
Unsharp measurements  are increasingly important for foundational insights in quantum theory and quantum information applications. Here, we report an experimental implementation of unsharp qubit measurements in a sequential communication protocol, based on a quantum random access code. The protocol involves three parties; the first party prepares a qubit system, the second party performs operations which return both a classical and quantum outcome, and the latter is measured by the third party. We demonstrate a nearly-optimal sequential quantum random access code that outperforms both the best possible classical protocol and any quantum protocol which utilises only projective measurements. Furthermore, while only assuming that the involved devices operate on qubits and that detected events constitute a fair sample, we demonstrate the noise-robust characterisation of unsharp measurements based on the sequential quantum random access code. We apply this characterisation towards quantifying the degree of incompatibility of two sequential pairs of quantum measurements.
\end{abstract}


\maketitle


\textit{Introduction.---} Textbook measurements in quantum theory are represented by complete sets of orthogonal projectors. However, general measurements in quantum theory are described by positive operator-valued measures (POVMs), i.e.~an ordered set of positive operators  $\{M_i\}_i$ with normalisation  $\sum_i M_i=\openone$. Evidently, projective measurements are instances of POVMs but not all POVMs are projective measurements. These non-projective measurements are well-defined in Hilbert spaces of fixed dimension (otherwise they can be viewed as projective measurements in a larger space \cite{NielsenChuang}). They are foundationally interesting and relevant to many phenomena and applications of quantum theory.

Some non-projective measurements are extremal in the space of all POVMs with fixed Hilbert space dimension and number of outcomes i.e.~they cannot be simulated with stochastic implementation of other measurements  \cite{Ariano}. Whereas such POVMs have been studied in broad contexts \cite{Ariano, Barnett, Derka, Brask, Renes2004, RenesAgain, TavakoliSIC, Gomez, Oszmaniec, tavakoliSmania}, far from all non-projective measurements are of this type. In fact, many interesting POVMs are \textit{unsharp} measurements, in the sense that they are weaker (noisy) variants of projective measurements. By suitably tuning the noise parameter (sharpness), an experimenter can control the information-disturbance trade-off \cite{Silva}; continuously from extracting no information and inducing no disturbance (non-interactive measurement) to extracting maximal information and inducing maximal disturbance (sharp projective measurement). Sequential unsharp measurements that individually induce only a small disturbance can be used for real-time monitoring of the evolution of single quantum systems \cite{Audretsch, Konrad, Korotkov2, Siddiqi}. When sufficiently frequent, such sequences effectively constitute continuous measurements, which have broad relevance in quantum information science (see e.g.~the review \cite{Clerk}). Two key application of sequential unsharp measurements are adaptive measurement protocols \cite{Armen, Lundeen} and quantum feedback protocols \cite{Cook, Gillett, Haroche}. Interestingly, such sequences are also versatile as they can be used to realise the most general quantum measurements \cite{Oreshkov}.  Moreover, unsharp measurements have prominent roles in a number of other topics including weak values \cite{Aharonov}, entanglement amplification \cite{Ota}, quantum random number generation \cite{Curchod}, tests of the memory-capacity of classical systems \cite{tavakoliCabello} and sequential quantum correlations \cite{Silva, Bera, Shenoy, Anwer, Korotkov}. This has prompted a number of experiments focused on the implementation of incompatible measurements \cite{Piacentini, Kim, Chen}, quantum contextuality \cite{Anwer} and quantum nonlocality \cite{Schiavon, Hu, Foletto}.

 \begin{figure}[t!]
 	\centering
 	\includegraphics[width=\columnwidth]{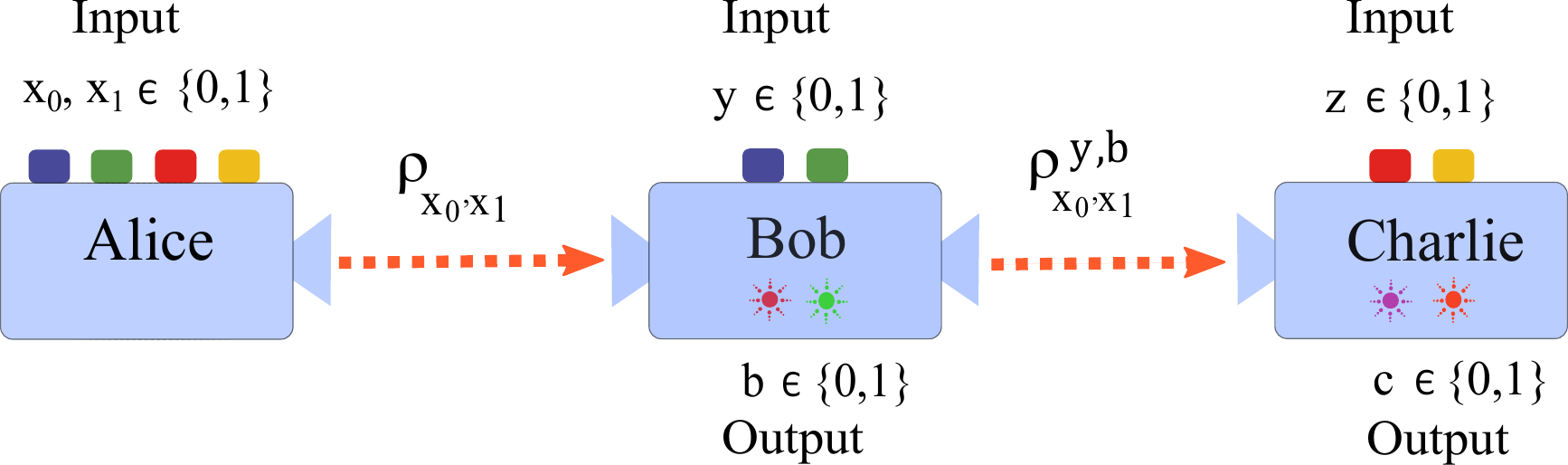}
 	\caption{Alice receives two bits $x_0,x_1$ and sends the qubit state $\rho_{x_0,x_1}$ to Bob who receives an input $y$ and produces a classical output $b$ and a quantum output $\rho_{x_0,x_1}^{y,b}$ which is measured by Charlie according to his input $z$, yielding an outcome $c$.}
 	\label{FigScenario}
 \end{figure}

Recently, Refs~\cite{Miklin, Mohan} considered unsharp measurements in a sequential implementation of a frequently encountered communication task known as a quantum random access code (QRAC) \cite{AmbainisNayak, AmbainisLeung, TavakoliHameedi}. In a (Q)RAC, a sender, Alice, receives two input bits $(x_0,x_1)$ which she encodes into a (qu)bit that is sent to a receiver, Bob. Bob receives an input bit $y$ and then attempts to choose his output $b$ such that it equals to Alice's $y$'th bit, i.e.~$b=x_y$. In an optimal classical protocol, Alice always sends $x_0$; thus Bob succeeds when $y=0$ and takes a random guess when $y=1$, leading to an average success probability of $0.75$. However, a quantum advantage is obtained if Alice prepares four qubit states forming a square on the equator of the Bloch sphere and Bob measures two suitably aligned Pauli observables, resulting in a success rate of $\approx 0.85$. From an alternative perspective, a QRAC can be viewed as a certification tool that allows an experimenter to characterise the involved preparation and measurement devices solely from its success rate, while assuming only that the setup operates on qubits \cite{tavakoliKaniewski}.

However, unsharp measurements in standard QRACs are unremarkable as their outcome statistics can be simulated by a measurement device that stochastically implements projective measurements. Therefore, Refs.~\cite{Miklin, Mohan} considered a sequential scenario (see Figure~\ref{FigScenario}) in which the post-measurement state of Bob is relayed to another receiver, Charlie, who receives an input bit $z$ and analogously attempts to recover the $z$'th bit of Alice. Thus, Alice sequentially implements a QRAC with  Bob and Charlie in the respective order. Here, unsharp measurements become indispensable: in order for both QRACs to achieve a high success rate, Bob must interact with the incoming system in such a way that sufficient information is extracted to power his guess of $x_y$, while simultaneously the disturbance is limited to allow Charlie to accurately guess $x_z$. Furthermore, it was shown  \cite{Miklin, Mohan} that sequential QRACs can serve as certification tools for characterising the unsharpness of  Bob's operations while only assuming that the states are qubits.

In this Letter, we report experimental implementation of sequential QRACs using measurements of tunable unsharpness and demonstrate nearly-optimal quantum correlations that outperform both all classical protocols as well as all quantum protocols based only on projective qubit measurements. We harvest these quantum communication advantages to certify the unsharpness parameter by confining it to a narrow interval. Subsequently, we theoretically develop and experimentally demonstrate how the sequential QRACs can be applied to quantify the degree of incompatibility \cite{Busch} between two sequential pairs of quantum measurements.

\textit{Scenario and theoretical background.---} Based on Refs.~\cite{Miklin, Mohan}, we describe the sequential QRAC experiment. It involves three parties, Alice, Bob and Charlie (see Figure~\ref{FigScenario}). Alice receives an input $x\equiv x_0,x_1\in\{0,1\}$ and prepares some uncharacterised qubit state denoted  $\rho_x$, which she sends to Bob. Bob receives an input $y\in\{0,1\}$ and performs a corresponding operation on $\rho_x$. This operation produces a classical output $b\in\{0,1\}$ and some post-operation qubit state denoted $\rho_x^{y,b}$, which is sent to Charlie. Charlie receives an input $z\in\{0,1\}$ and then measures $\rho_x^{y,b}$, yielding an outcome $c\in\{0,1\}$. All inputs $(x,y,z)$ are statistically independent and uniformly distributed. The limit of many rounds yields conditional probability distributions $p(b,c|x,y,z)$.

The conditional probability distributions $p(b,c|x,y,z)$ are used to evaluate the success rate of two QRACs: one between Alice and Bob, and one between Alice and Charlie. The former is successful when $b=x_y$ and the latter is successful when $c=x_z$. The two respective success rates read
\begin{align}\nonumber
& W_\text{AB}=\frac{1}{8}\sum_{x,y}P(b=x_y|x,y),\\\label{witness}
& W_{\text{AC}} = \frac{1}{8}\sum_{x,z}P(c=x_z|x,z).
\end{align}
Note that we can always take $W_\text{AB},W_\text{AC}\in[\frac{1}{2},1]$. Evidently, $W_\text{AB}$ is independent of Charlie. However, $W_\text{AC}$ is not independent of Bob because he operates on the system before it reaches Charlie. 

Bob's two operations ($y=0,1$) are described by the notion of a quantum instrument \cite{Teiko}, which captures both the measurement statistics and the evolution of the measured state. A quantum instrument is defined as an ordered set of trace-non-increasing completely positive maps $\{\Lambda_{b|y}\}_b$ with the property that for any state $\rho$ it holds that $p(b|y)=\Tr \left(\Lambda_{b|y}(\rho)\right)$. Having observed the classical output $b$, the quantum output of the instrument is $\rho^{y,b}=\Lambda_{b|y}(\rho)/\Tr\left(\Lambda_{b|y}(\rho)\right)$. Since we consider qubits and Bob has binary outcomes, the extremal quantum instruments are  written as $\Lambda_{b|y}(\rho)=K_{b|y}\rho K_{b|y}^\dagger$, where $\{K_{b|y}\}_b$ are Kraus operators  satisfying $\sum_{b}{K_{b|y}}^\dagger K_{b|y} = \openone$, with the convenient property that ${K_{b|y}}^\dagger K_{b|y}=B_{b|y}$ where $\{B_{b|y}\}_b$ are the two POVMs of Bob \cite{Pellonpaa}. For simplicity, we can represent Bob's measurements in terms of two observables which, in general, read $B_{y}\equiv B_{0|y}-B_{1|y}=\alpha_y \openone+\vec{n}_y\cdot \vec{\sigma}$, where $\vec{n}_y$ are Bloch vectors, $\vec{\sigma}$ are the Pauli matrices and $|\alpha_y|\leq 1-|\vec{n}_y|$. The sharpness of Bob's measurements is defined as $\eta_y=|\vec{n}_y|$. Notice that for $\eta_y\in\{0,1\}$, the measurements are non-interactive and sharp respectively, whereas $\eta_y\in(0,1)$ corresponds to  intermediate cases. We consider the case of $\eta\equiv \eta_0=\eta_1$. We emphasise that one can stochastically simulate Bob's unsharp POVMs using only projective measurements, but one cannot simulate his quantum instrument in this manner. Therefore, we can distinguish a projective simulation from a genuine unsharp measurement by considering both the classical and quantum output.

By inspecting the witnesses $(W_\text{AB},W_\text{AC})$, one may characterise the sharpness parameter $\eta$. Refs.~\cite{Miklin, Mohan} showed that for a given value of $W_\text{AB}$, the optimal value of $W_\text{AC}$ in quantum theory is given by
\begin{equation}\label{qopt}
W_\text{AC}=\frac{1}{8}\left(4+\sqrt{2}+\sqrt{16W_\text{AB}-16W_\text{AB}^2-2}\right),
\end{equation}
and that such an optimal pair implies a precise value of $\eta$. However, in the experimentally realistic case in which perfectly optimal quantum correlations are not relevant, a sub-optimal witness pair can be used to deduce upper and lower bounds on $\eta$,
\begin{align}\nonumber
&  \eta \geq\sqrt{2}\left(2W_\text{AB}-1\right)\equiv \eta_\text{min},\\ \label{eq4}
& \eta\leq 2\sqrt{\left(2+\sqrt{2}-4W_\text{AC}\right)\left(2W_\text{AC}-1\right)}\equiv \eta_\text{max}.
\end{align}
Thus, the closer the experimentally observed correlations are to the optimal ones in Eq.~\eqref{qopt}, the narrower is the interval $I(W_\text{AB},W_\text{AC})\equiv [\eta_\text{min},\eta_\text{max}]$ to which we can confine the sharpness $\eta$.

\textit{Experiment.---} The optimal quantum correlations \eqref{qopt} are obtained with a unique quantum strategy (up to a global unitary) \cite{Mohan}. Alice needs to prepare four states forming a square on a great circle on the Bloch sphere. For simplicity we choose the $xz$-plane and Alice's four states $|\psi_{x_0x_1} \rangle = \cos\alpha_{x_0x_1} \hspace{1mm}|0\rangle + \sin\alpha_{x_0x_1} \hspace{1mm}|1\rangle $  corresponding to the four values  $\{\frac{\pi}{8}$,$-\frac{3\pi}{8}$,$\frac{9\pi}{8},\frac{5\pi}{8}\}$ of $\alpha_{x_0x_1}$ respectively, where $\rho_x=\ketbra{\psi_{x_0x_1}}{\psi_{x_0x_1}}$. Similarly, the optimal quantum instruments of Bob correspond to the Kraus operators $K_{b|y}=\sqrt{\left(\openone+ (-1)^b  B_y\right)/2}$  for a suitably chosen $\eta$, where $B_y\in\{\eta\sigma_x,\eta\sigma_z\}$ are the corresponding observables of Bob. The quantum output is sent to Charlie whose observables are two complementary projective measurements $C_0=\sigma_{x}$ and $C_1=\sigma_{z}$. In an ideal experiment, for every $\eta$, we obtain the witness pair,
\ba
W_{\text{AB}} = \frac{2 + \sqrt{2} \eta}{4}, \hspace{5mm}
W_{\text{AC}} = \frac{4 + \sqrt{2} + \sqrt{2-2\eta^2}}{8},
\label{EqExp1}
\ea
which satisfies the optimality condition  \eqref{qopt}.

\begin{figure}[t]
	\centering
	\includegraphics[width=\columnwidth]{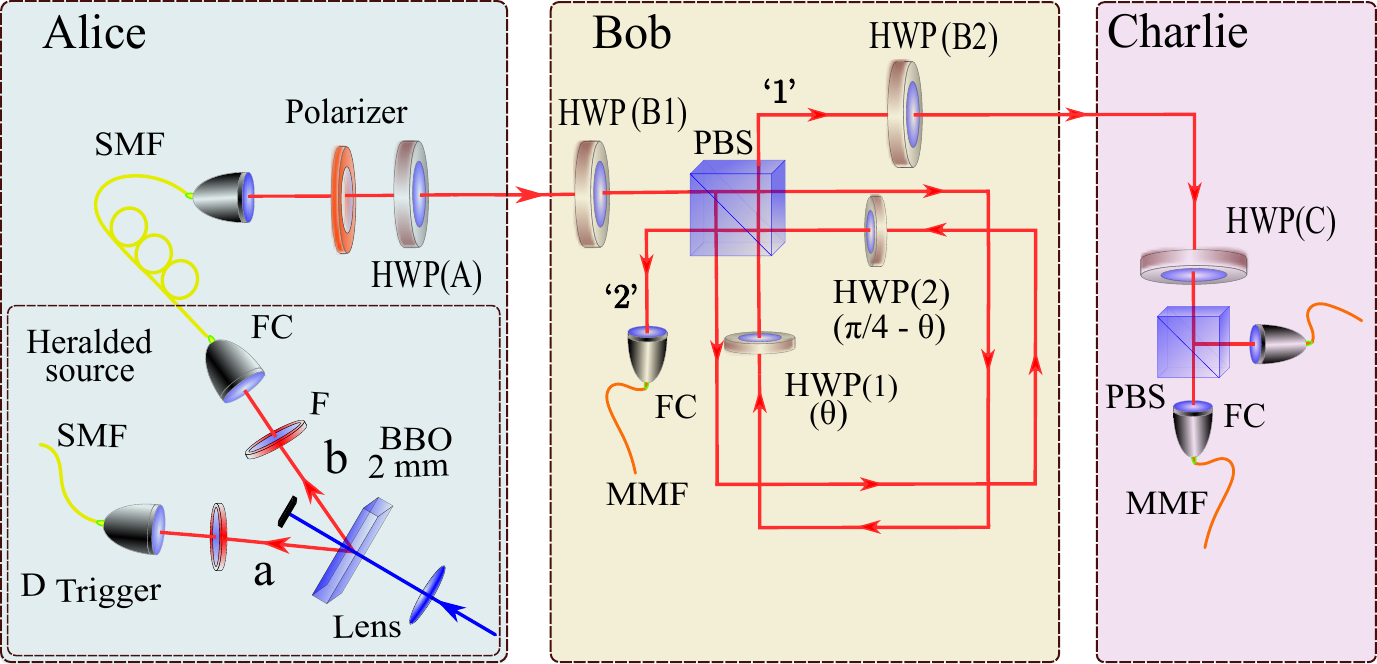}
	\caption{Experimental setup. Alice prepares her states using a heralded photon source, a polariser and a half-wave plate HWP(A). Bob's instrument is realised by a shifted Sagnac interferometer where the sharpness parameter $\eta=\cos(4 \theta$) is tuned using half-wave plates HWP(1) and HWP(2). HWP(B1) and HWP(B2) are used to switch between the observables $B_0$ and $B_1$ as well as selecting the output corresponding to the outcome $b=0$ and $b=1$. Charlie performs projective (sharp) measurements on the received qubit from Bob using a HWP(C) and a polarisation beam splitter (PBS).}
	\label{figure2}
\end{figure}

We implemented this optimal strategy, using single-photon polarisation qubits where the computational basis corresponds to horizontal (H) and vertical (V)
polarisation, i.e. $\ket H \equiv \ket 0$ and  $\ket V \equiv \ket 1$. The complete optical setup is shown in Figure~\ref{figure2}. Alice's preparation device also encloses a heralded single photons source that produces photons at wave-length 780 nm through spontaneous parametric down conversion (SPDC) by pumping a type-I beta barium borate (BBO) single crystal of thickness 2 mm using 390 nm femto-second laser pulses. Time correlated idler and signal photons are  spectrally and spatially purified by passing through 3 nm (FWHM) wide optical filters (F) and coupling into single mode fibers (SMF) respectively. The idler photons in mode `a' are detected by an avalanche photo-diode (APD), marked as $\text{D}_{\text{Trigger}}$, with detection efficiency $\sim 60\%$, which produces a trigger signal indicating the presence of a photon in mode `b'. Alice prepares this photon in one of the four desired states $\ket{\psi_{x_0,x_1}}$  using a polariser when it only passes through $\ket H $ and a half-wave plate, HWP(A), at angles $11.25^\circ$, $-11.25^\circ$ $33.75^\circ$ and $-33.75^\circ$ respectively and sends it to Bob. 

Bob's unsharp measurements on the received photons are performed using shifted Sagnac interferometer as described in \cite{Anwer,Hu}. In this setup the strength of the measurement is controlled by rotating half-wave plate HWP(1) to $\theta$ and HWP(2) to $\frac{\pi}{4} - \theta$, that are placed respectively in the path of horizontally and vertically polarised beams after the polarisation beam splitter (PBS) such that $\eta = \cos (4 \theta)$. To switch between the bases $B_y$ according to the input $y$, Bob rotates both his wave-plates HWP(B1) and HWP(B2) to $22.5^\circ$ and $0^\circ$ respectively. The outcome of these measurements $b \in \{0,1\}$ is encoded in the output path of the interferometer such that $b = 0$ $(b = 1)$ corresponds to the detection of the photon in the output path `1' $\equiv$ transmission (`2' $\equiv$ reflection). In a sequential scenario, we choose to consider only one output path at a time to simplify the setup and by adding an additional rotation to the HWP(B1) and HWP(B2), we can select the output we want to analyse at a given time. Using output `2', Bob will locally be able to learn the outcome of his measurement counterfactually when using perfect detectors. Also, when the fair sampling assumption is invoked, which is the case in this experiment, Bob can still infer his outcome of the measurement locally using average photon rates.

Finally, Charlie's projective measurement setup is consisted of HWP(C), PBS, a pair of fiber couplers (FC) and multi-mode fibers (MMF) that propagate the photons to a pair of APDs. He performes $C_z \in \{ \sigma_{x}, \sigma_{z} \}$ on the received qubits according to his random input $z \in \{0,1 \}$, by rotating HWP(C) to $22.5^\circ$ and $0^\circ$ respectively. The results of Charlie's marginal probabilities (for evaluating $W_\text{AC}$) are obtained by averaging out the inputs and outputs of Bob. 

\begin{figure}[t]
	\centering
	\includegraphics[width=\columnwidth]{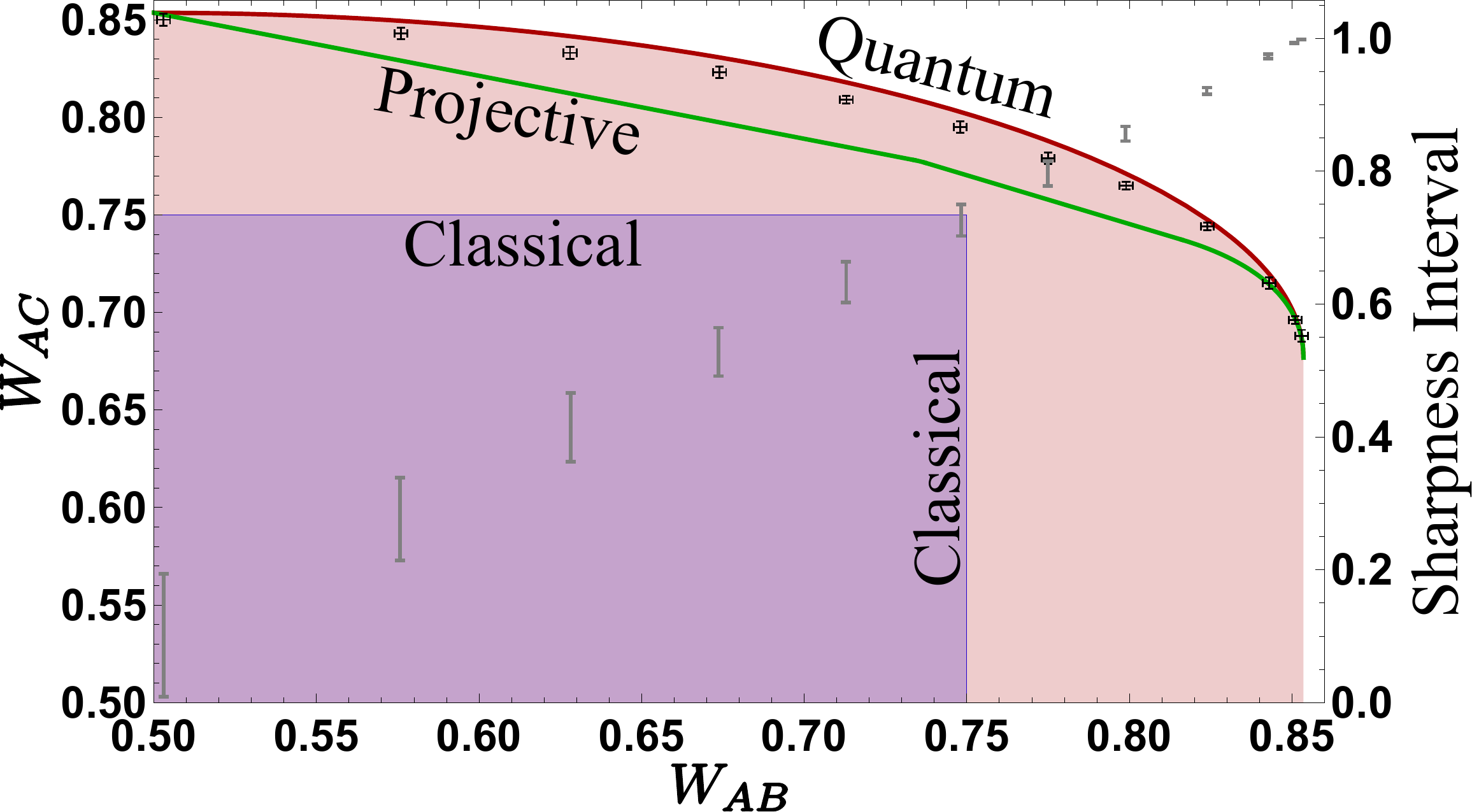}
	\caption{Experimental results. Optimal quantum correlations (Red), optimal quantum correlations from stochastic projective measurements (Green), optimal classical correlations (Blue) and experimentally obtained values of witness pairs $(W_{\text{AB}},W_{\text{AC}})$ (Black). The characterisation of the sharpness parameter $\eta$ is depicted by gray bars corresponding to the interval to which it is confined (y-axis on the right-hand-side).}
	\label{figure3}
\end{figure}

\textit{Results.---} To evaluate $(W_{\text{AB}},W_{\text{AC}})$ from the data, we require the marginal probabilities appearing in Eq~\eqref{witness}. All parties setting are set using motorised rotation stages that are  controlled by a computer program. To gather sufficient statistics we measure $60$ sec in each setting  with a rate of $\sim$ $20$ kHz and collected at least $1.2$ million events. Each measured value of $(W_{\text{AB}},W_{\text{AC}})$ together with the (black colour) error bars (horizontal and vertical corresponding  to $W_{\text{AB}}$ and $W_{\text{AC}}$ respectively) is shown in Figure~\ref{figure3} and can be compared to the optimal quantum correlations (red colour) and the optimal classical correlations (blue colour, given by $(W_\text{AB},W_\text{AC})\leq 3/4$). Our obtained quantum correlations are nearly optimal for all considered values of $\eta$. Also, in the worst case, the classical limit is outperformed by at least $15$ standard deviations. Moreover, the data reliably outperforms the optimal quantum correlations attainable when Bob uses stochastic projective measurements (green colour) (see Ref.~\cite{Miklin}). This certifies the advantage of unsharp measurements in sequential QRACs. Notably, the projective bound is not outperfromed for the two data points corresponding to  $\eta\in\{0,1\}$ since in these cases the   bound coincides with the optimal quantum correlations.

From the inequalities in Eq.~\eqref{eq4}, we can determine an upper and a lower bound on the sharpness parameter. Thus $\eta$ can be confined to the interval $I(W_\text{AB},W_\text{AC})$ for each of the measured values of the witness pair $(W_\text{AB},W_\text{AC})$. These certified intervals are  depicted by gray bars in Figure~\ref{figure3} located vertically from the corresponding witness' and using the y-axis on the right side. We observe that the certification is more precise (the interval is smaller) as the sharpness parameter increases. The smallest (largest) interval, corresponding to an essentially projective (non-interactive) measurement, has a width of about $10^{-3}$ ($0.2$). This is due to the fact that the bounds in Eq~\eqref{eq4} become more sensitive to small imperfections when $W_\text{AC}$ increases. Further details about the experimental data is presented in Appendix~\ref{AppData}. Moreover, in Appendix~\ref{AppTomography}, we also compare this characterisation of unsharp measurements to a simple tomographic model with an essentially trusted preparation device subjected to comparably small errors.

\textit{Data Analysis.---} The experiment is influenced by systematic errors originating from, for instance, imperfect wave-plates as well as offsets in their marked zero position, finite PBS extinction and cross-talk, and limited interference visibility. The magnitude of these errors is revealed by the extent to which the experimental points are shifted away from the optimal quantum correlations. In order to minimise systematic errors, we carefully select and characterise all the optical components. This brings us closer to the optimal quantum correlation and the experimental points correspond to a more than $98\%$ total visibility estimation. Nevertheless, random errors due to Poissonian statistics or due to repetition of the experimental settings with limited precision will spread the observed point on the Figure~\ref{figure3} to a region contained within the black bars. To keep this error low, all the settings are set by computerised controlled motors with repetition precision $<0.02^\circ$. Errors together with mean values are provided in Appendix~\ref{AppData}.

\begin{figure}
	\centering
	\includegraphics[width=0.95\columnwidth]{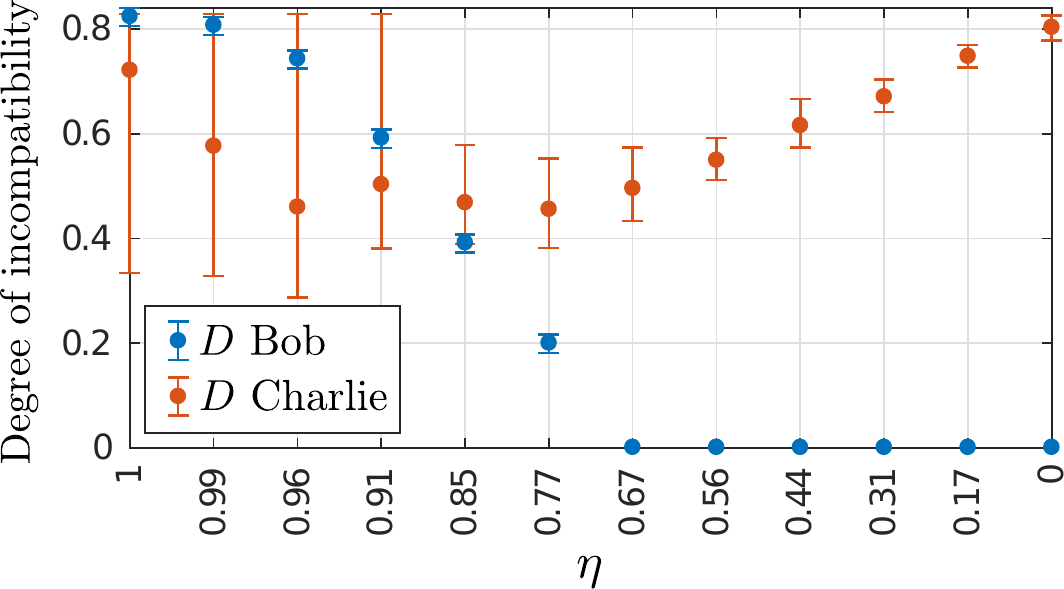}
	\caption{Lower bound on the degree of incompatibility in Bob's (blue) and Charlie's (orange) respective pair of measurements for the twelve different targeted values of the sharpness parameter $\eta$.}\label{FigDOI}
\end{figure}

\textit{Quantifying sequentual measurement incompatibility.---} In order to witness quantum correlations, one requires incompatible measurements. In that sense, violating the classical constraint with $W_\text{AB}$ ($W_\text{AC}$) certifies that Bob's (Charlie's) two POVMs are incompatible \cite{Carmeli, Uola}. It is, however, more informative to consider a quantitative inference; is it possible to deduce from $(W_\text{AB},W_\text{AC})$ a lower bound on the extent to which Bob's and Charlie's POVMs are incompatible? In order to achieve such quantification of Heisenberg uncertainty, one must first define a measure of incompatibility valid for dichotomic qubit observables. We use the \textit{degree of incompatibility} introduced in Ref.~\cite{Busch}; 
\begin{equation}\label{DOI}
D(\vec{n}_0,\vec{n}_1)=|\vec{n}_0+\vec{n}_1|+|\vec{n}_0-\vec{n}_1|-2,
\end{equation}
where $\vec{n}_0$ and $\vec{n}_1$ are the Bloch vectors of the observables. All compatible observables obey $D\leq 0$ whereas incompatible observables obey $D \leq 2\left(\sqrt{2}-1\right)$.  As expected, the bound is saturated by two Pauli observables. Since we are interested in incompatible observables, we simply reset negative values of $D$ to $0$. As shown in Appendix~\ref{AppIncompatibility}, the success rate of a QRAC implies a lower bound on $D$:
\begin{equation}\label{b1}
D\geq 8W-6.
\end{equation}	
Thus, whenever a QRAC exceeds the classical bound of $\frac{3}{4}$, a degree of incompatibility is certified and quantified. By choosing $W=W_\text{AB}$, we use Eq~\eqref{b1} to quantify the incompatibility of Bob's unsharp measurements. The bound in Eq.~\eqref{b1} can also be applied to Charlie's measurements, but it would significantly underestimate their degree of incompatibility due to the sequential nature of the experiment. A more sophisticated quantification is possible when exploiting both $W_{AB}$ and $W_{AC}$ and the fact that $\eta\in I(W_\text{AB},W_\text{AC})$. Considering unbiased observables for Bob, i.e.~$B_y=\eta (\hat{n}_y\cdot \vec{\sigma})$, where $\hat{n}_y$ is the normalised Bloch vector, we show in Appendix~\ref{AppIncompatibility} that Charlie's degree of incompatibility respects
\begin{align}\label{b2}
& D\geq \min_{\eta\in I(W_\text{AB},W_\text{AC})}\frac{16W_\text{AC}-8}{1+g_\eta+f_{W_\text{AB}}\left(1-g_\eta\right)}-2
\end{align}
where $g_\eta\equiv\sqrt{1-\eta^2}$ and $
f_{W_\text{AB}}\equiv 2\frac{\eta_{\text{min}}}{\eta}\sqrt{1-\left(\frac{\eta_{\text{min}}}{\eta}\right)^2}$.

Notice that if we choose not to exploit the certified interval $I(W_\text{AB},W_\text{AC})$, we may simply take the limit of $\eta\rightarrow 0$ and recover the bound in Eq~\eqref{b1} for $W=W_\text{AC}$. In Figure~\ref{FigDOI} we show the degree of incompatibility as obtained from the twelve experimentally measured witness pairs $(W_\text{AB},W_\text{AC})$ corresponding to different targeted values of the sharpness parameter $\eta$. As expected, we see that the incompatibility of Bob's measurements decreases with $\eta$ and vanishes in the vicinity of $\eta=1/\sqrt{2}$, which is the theoretical threshold. For Charlie, we always find a high degree of incompatibility which stems from his projective measurements.

\textit{Conclusions.---}  By precise control of unsharp quantum measurements, we demonstrated nearly optimal sequential quantum random access codes that outperform not only the best possible classical protocols but also the best possible quantum protocols based only on projective measurements. We harvested the quantum advantage in the communication task in order to certify the degree of unsharpness in the preformed measurements. Exploiting both the sequential QRACs and the certification of the unsharpness, we quantitatively demonstrated the incompatibility of two sequential pairs of measurements accross a wide range of sharpness parameters. Our results demonstrate the usefulness of unsharp measurements in quantum communication tasks, the possibility of quantifying the degree of incompatibility of sequential pairs of unsharp observables and the practical feasibility of characterising them under weak assumptions.

\section*{Acknowledgements}
This work was supported by the Swedish research council, Knut and Alice Wallenberg Foundation and the Swiss National Science Foundation (Starting grant DIAQ, NCCR-QSIT). NM acknowledges the financial support by the First TEAM Grant No. 2016-1/5.

\appendix
\onecolumngrid
\section{Experimental data}\label{AppData}

\begin{table*}[ht]
	\centering
	\caption{\label{Table_3_1} Details of the results presented in the main text, showing half-wave plates rotation $\theta$ used to tune the sharpness $\eta$ of the Bob's effective measurement, its corresponding values,  witness pair $(W_{\text{AB}},W_{\textbf{AC}})$, certified sharpness (upper and lower bound) and sharpness interval together with their errors.}
	\renewcommand{\arraystretch}{1.2}
	\begin{tabular}{>{\centering}m{1in} >{\centering}m{1in} >{\centering}m{0.9in} >{\centering}m{1in} >{\centering}m{1in} >{\centering\arraybackslash}m{1in} >{\centering\arraybackslash}m{1in}}
		\toprule
		\bf Half-wave-plate angle $\boldsymbol{\theta}$ & \bf Measurement Sharpness $\mathbf{\eta}=\mathbf{cos(4 \boldsymbol{\theta})}$ & $\mathbf{W_{\text{AB}}}$ & $\mathbf{W_{\text{AC}}}$ & \begin{tabular}{ll} \multicolumn{2}{c}{\bf Sharpness Bound} \\[0.5em]\midrule \bf {Lower Bound} & \bf{Upper Bound} \end{tabular} & & \bf {Interval} $\mathbf{(\Delta)}$ \\[0.5em]
		\midrule
		0 & 1.000 & 0.853 $\pm$ 0.002 & 0.688 $\pm$ 0.003 & 0.998 $\pm$ 0.006 & 1.00 $\pm$ 0.01 & 0.00 $\pm$ 0.01 \\
		2 & 0.990 & 0.851 $\pm$ 0.002 & 0.696 $\pm$ 0.002 & 0.992 $\pm$ 0.006 & 0.994 $\pm$ 0.008 & 0.00 $\pm$ 0.01 \\
		4 & 0.961 & 0.843  $\pm$ 0.002 & 0.715  $\pm$ 0.003 & 0.969 $\pm$ 0.006 & 0.98 $\pm$ 0.01 & 0.01 $\pm$ 0.01 \\
		6 & 0.914 & 0.824  $\pm$ 0.002 & 0.744 $\pm$ 0.002 &  0.916 $\pm$ 0.006 & 0.93 $\pm$ 0.01 & 0.01 $\pm$ 0.01 \\
		8 & 0.848 & 0.799  $\pm$ 0.002 & 0.765 $\pm$ 0.002 &  0.845 $\pm$ 0.006 & 0.87 $\pm$ 0.01 & 0.02 $\pm$ 0.01 \\
		10 & 0.766 & 0.775  $\pm$ 0.002 & 0.779   $\pm$ 0.003 &  0.778 $\pm$ 0.006 & 0.82 $\pm$ 0.02 & 0.04 $\pm$ 0.02 \\
		12 & 0.669 & 0.748  $\pm$ 0.002 & 0.795  $\pm$ 0.003 & 0.702 $\pm$ 0.006 & 0.75 $\pm$ 0.02 & 0.05 $\pm$ 0.02 \\
		14 & 0.559 & 0.713  $\pm$ 0.002 & 0.809  $\pm$ 0.002 &  0.602 $\pm$ 0.006 & 0.66 $\pm$ 0.02 & 0.06 $\pm$ 0.02 \\
		16 & 0.438 & 0.674  $\pm$ 0.002 & 0.823  $\pm$ 0.003 &  0.491 $\pm$ 0.006 & 0.56 $\pm$ 0.03 & 0.07 $\pm$ 0.03 \\ 
		18 & 0.309 & 0,628  $\pm$ 0.002 & 0.833  $\pm$ 0.003 &  0.363 $\pm$ 0.006 & 0.47 $\pm$ 0.03 & 0.10 $\pm$ 0.03 \\
		20 & 0.174 & 0.576  $\pm$ 0.002 & 0.843  $\pm$ 0.003 & 0.214 $\pm$ 0.006 & 0.34 $\pm$ 0.05 & 0.13 $\pm$ 0.05 \\
		22.5 & 0.000 & 0.503  $\pm$ 0.002 & 0.850  $\pm$ 0.003 & 0.009 $\pm$ 0.006 & 0.20 $\pm$ 0.01 & 0.2 $\pm$ 0.1 \\
		\bottomrule
	\end{tabular}
\end{table*}

\section{Comparison to error-bounded detector tomography}\label{AppTomography}

There are many methods of performing qubit tomography. We focus on a particularly simple case of scaled direct inversion tomography for an unbiased qubit observable. We need to introduce a preparation device which is fully controlled, i.e.~it is assumed to flawlessly prepare states. In general, one requires a tomographically complete set of preparations. A simple and unbiased choice of such preparations is four states that form a tetrahedron on the Bloch sphere. The corresponding Bloch vectors can be taken as 
\begin{align}\label{tetra}\nonumber
& \vec{a}_1=\frac{(1,1,1)}{\sqrt{3}} & \vec{a}_2=\frac{(1,-1,-1)}{\sqrt{3}}\\
& \vec{a}_3=\frac{(-1,1,-1)}{\sqrt{3}} & \vec{a}_4=\frac{(-1,-1,1)}{\sqrt{3}},
\end{align}
and the states are denoted $\rho^\text{ideal}_x=1/2\left(\openone+\vec{a}_x\cdot\vec{\sigma}\right)$.  Since a general unbiased qubit observable reads $E^\text{lab}=\vec{n}\cdot \vec{\sigma}$, the probabilities $p_x=\Tr\left(\rho_x^\text{ideal} E^\text{lab}\right)$ can be written as 
\begin{equation}
\vec{p}=A\vec{n}
\end{equation}
where $A$ is the $4\times 3$ matrix with rows $\vec{a}_1,\vec{a}_2,\vec{a}_3,\vec{a}_4$. Hence, given the observation $\vec{p}$, we can deduce the initially unknown observable by evaluating
\begin{equation}\label{est}
\vec{n}=(A^\text{T}A)^{-1}A^\text{T}\vec{p}.
\end{equation}

However, no realistic experiment can live up to the idealisation of fully controlled preparation devices. To make a more realistic model, we introduce an error parameter $\epsilon\in[0,1]$ for the preparation device. We consider that the preparation device outputs the desired states \eqref{tetra} with average fidelity
\begin{equation}\label{fid}
\frac{1}{4}\sum_{x=1}^{4} F(\rho^\text{ideal}_x,\rho^{\text{lab}}_x)\geq 1-\epsilon,
\end{equation}
where $\rho^\text{lab}_x$ are the de-facto states prepared in the experiment. Notably, since the target states are pure, we have that $F(\rho^\text{ideal}_x,\rho^{\text{lab}}_x)=\Tr\left(\rho^\text{ideal}_x\rho^\text{lab}_x\right)$. Thus, the smaller we choose $\epsilon$, the closer must the laboratory states be to the targeted (tetrahedron) states. Notice that for $\epsilon=0$, we must have $\rho^\text{lab}_x=\rho^\text{ideal}_x$. The introduced error means that the observed probabilities now read $p_x=\Tr\left(\rho_x^\text{lab}E^\text{lab}\right)$ which via Eq.~\eqref{est} will lead to a somewhat inaccurate estimate ($\vec{n}^\text{est}$, corresponding to $E^\text{est}$) of the de-facto observable ($E^\text{lab}$, with Bloch vector $\vec{n}^\text{lab}$). It may happen that $\vec{n}^\text{est}$ is not a valid Bloch vector; in such cases we renormalise it by letting $\vec{n}^\text{est}\rightarrow \vec{n}^\text{est}/|\vec{n}^\text{est}|$.

\begin{figure}[h!]
	\centering
	\includegraphics[width=0.5\columnwidth]{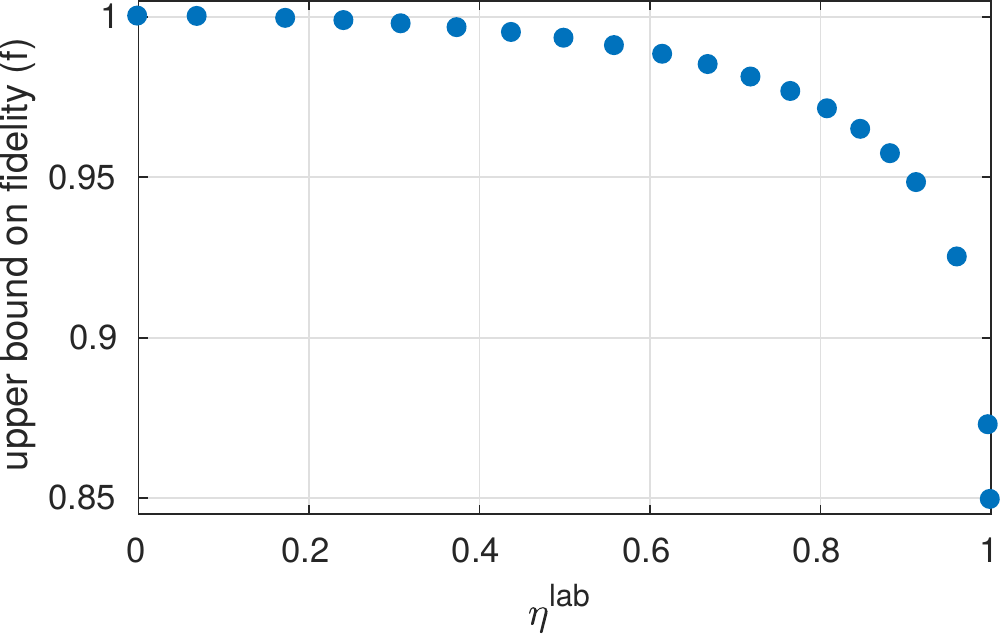}
	\caption{Upper bound the fidelity $f(\epsilon,E^\text{lab})$ for $E^\text{lab}=\eta^\text{lab}\sigma_\text{z}$  in the detector tomography scheme using four tetrahedral states subject to $\epsilon=1\%$ average error.}\label{FigureFidelity}
\end{figure}

To what extent can a given $\epsilon$ influence the accuracy of the tomography? We evaluate the accuracy of the tomography in terms of the fidelity between the outcome-zero operators of $E_0^\text{lab}=\left(\openone+E^\text{lab}\right)/2$ and $E_0^\text{est}=\left(\openone+E^\text{est}\right)/2$ \footnote{Notice that the zero-outcome operators are also valid states. Notably, choosing said operator fixes the full POVM, which is why we ignore the one-outcome operators in the fidelity.}. Then, it is natural to consider the worst estimation compatible with the allowed error. This amounts to solving
\begin{equation}\label{min}
f(\epsilon,E^\text{lab})=\min_{\{\rho^\text{lab}_x\}} F(E^\text{lab}_0,E^\text{est}_0),
\end{equation}
with $p_x=\Tr\left(\rho_x^\text{lab}E^\text{lab}\right)$, the estimation performed via Eq.~\eqref{est} and the laboratory states obeying the constraint in Eq.~\eqref{fid}. We remark that the objective can be written on the simple form
\begin{equation}
F(E^\text{lab}_0,E^\text{est}_0)=\Tr\left(E^\text{lab}_0E^\text{est}_0\right)+2\sqrt{\det(E^\text{lab}_0)\det(E^\text{est}_0)}.
\end{equation}

\begin{figure}[]
	\centering
	\includegraphics[width=0.5\columnwidth]{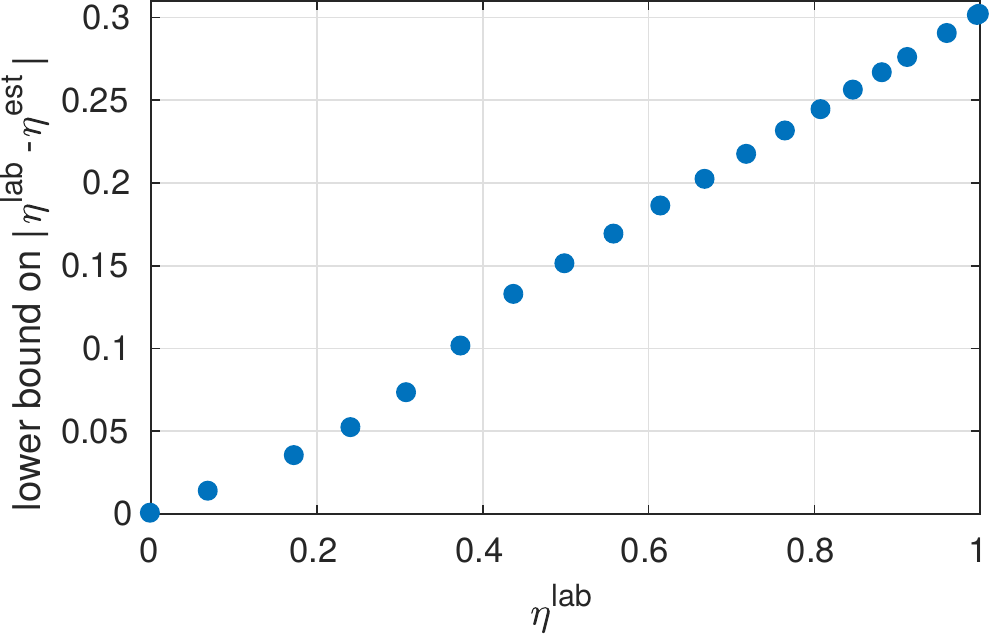}
	\caption{Lower bound on the imprecision of the estimated sharpness of the laboratory observable $E^\text{lab}=\eta^\text{lab}\sigma_\text{z}$ in the detector tomography scheme using four tetrahedral states subject to $\epsilon=1\%$ average error.}\label{FigError}
\end{figure}

A reasonable choice of error, on par with that estimated to be present in our experiment in the main text, is $\epsilon=1\%$. Fixing this error, we have numerically implemented the minimisation in Eq.~\eqref{min} for a laboratory observable of the form $E^\text{lab}=\eta^\text{lab}\sigma_\text{z}$. We performed the minimisation for several different values of $\eta$ and the results are displayed in Figure~\ref{FigureFidelity}. We see that as the observable becomes sharper, the impact of the error becomes greater. However, since the numerics are not guaranteed to find the global minimum, our results constitute upper bounds on  $f(\epsilon,E^\text{lab})$. Thus, we run the risk of overestimating the guaranteed accuracy of the detector tomography scheme.

For every choice of $\eta^\text{lab}$, our numerics return the estimate $E^\text{est}$. We compute the sharpness ($\eta^\text{est}$) of this observable and consider how far it is from that of the targeted value ($\eta^\text{lab}$), i.e.~we consider  $|\eta_\text{est}-\eta_\text{lab}|$. The results are shown in Figure~\ref{FigError}. In analogy with the fidelity, we find that the possible imprecision of the estimated sharpness parameter in the scheme increases with the sharpness of the observable. It is interesting to note that for a nearly-projective observable, an error of $\epsilon=1\%$ is compatible with an estimated observable whose sharpness is between $25\%-30\%$ incorrect.

\section{Quantifying the degree of incompatibility through a quantum random access codes}\label{AppIncompatibility}
In this section, we show how to use the measured values of the two QRACs ($W_\text{AB}$ and $W_\text{AC}$) to deduce lower bounds on the degree of incompatibility ($D$) of Bob's effective POVMs and Charlie's POVMs respectively. We first show how to connect QRACs to the degree of incompatibility, which can immediately be applied to Bob's effective POVMs. Then, we show how to take the decohering influence of Bob into account in order to also bound the incompatibility of Charlie's POVMs.

\subsection{Quantifying Bob's degree of incompatibility from a QRAC}
We use the Bloch sphere representation to write the qubit preparations as $\rho_x=\left(\openone+\vec{a}_x\cdot \vec{\sigma}\right)/2$ and the qubit observables as $B_y=\alpha_y\openone+\vec{n}_y\cdot \vec{\sigma}$, where $\vec{a}_x$ and $\vec{n}_y$ are Bloch vectors and $|\alpha_y|\leq 1-|\vec{n}_y|$. Then, we can write the QRAC as
\begin{equation}\label{Wit}
W=\frac{1}{2}+\frac{1}{16}\sum_{x,y} (-1)^{x_y}\Tr\left(\rho_xB_y\right)
=\frac{1}{2}+\frac{1}{8}\left[\vec{r}_0\cdot \left(\vec{n}_0+\vec{n}_1\right)+\vec{r}_1\cdot \left(\vec{n}_0-\vec{n}_1\right)\right],
\end{equation}
where we have defined
\begin{equation}\label{r}
\vec{r}_z=\frac{\vec{a}_{0z}-\vec{a}_{1\bar{z}}}{2},
\end{equation}
where $\bar{z}$ is the bitflip of $z$. Now, for a given pair of measurements, let us maximise the QRAC over the preparations of Alice. Evidently, her optimal choice is to align $\vec{r}_0$ with $\left(\vec{n}_0+\vec{n}_1\right)$ and $\vec{r}_1$ with $\left(\vec{n}_0-\vec{n}_1\right)$. This yields
\begin{equation}
\max_{\{\rho_x\}} W=\frac{1}{2}+\frac{1}{8}\left(|\vec{n}_0+\vec{n}_1|+|\vec{n}_0-\vec{n}_1|\right)
=\frac{1}{2}\left(1+\frac{D(\vec{n}_0,\vec{n}_1)+2}{4}\right),
\end{equation}
where $D$ is the degree of incompatibility. Hence, for some arbitrary observed success probability in a QRAC, it therefore holds that
\begin{equation}
W\leq \frac{1}{2}\left(1+\frac{D(\vec{n}_0,\vec{n}_1)+2}{4}\right),
\end{equation}
which is rearranged to
\begin{equation}\label{DOI}
D(\vec{n}_0,\vec{n}_1)\geq 8W-6.
\end{equation}
This certification of Heisenberg uncertainty through the QRAC naturally applies to Bob's effective POVMs in the sequential experiment since the Alice-Bob QRAC makes no reference to Charlie. Hence, we can immediately bound the incompatibility of Bob's pair of measurements via Eq~\eqref{DOI} by taking $W=W_\text{AB}$.

Naturally, one can also use Eq~\eqref{DOI} to bound the incompatibility of Charlie's pair of POVMs by taking $W=W_\text{AC}$. However, in the sequential scenario, this is typically a significant underestimate of Charlie's incompatibility. The reason is that Bob's influence decoheres the effective preparations that Charlie receives, thus lowering the reachable values of $W_\text{AC}$ even if his two measurements are maximally incompatible. The bound \eqref{DOI} does not take the disturbance of Bob into account if we set $W=W_\text{AC}$. Therefore, we derive a more sophisticated quantification of the incompatibility of Charlie's measurements based on the experimental confinement of Bob's sharpness parameter $\eta$ to a restricted interval.

\subsection{Quantifying Charlie's degree of incompatibility from two QRACs}
The quantum instrument of Bob decoheres the preparations of Alice such that the effective preparations reaching Charlie contain less information about her input.  This typically corresponds to a reduction in the purity of the original preparations. Therefore, with the aim of deriving a better bound than Eq~\eqref{DOI} for quantifying also the incompatibility of Charlie's POVMs, we  first address the purity of Bob's post-measurement states.

Let $\hat{n}_y$ be the directions of Bloch vectors (i.e. normalized Bloch vectors) of Bob's measurements as before, i.e. $B_y = \eta\hat{n}_y\cdot\vec{\sigma}$ are Bob's observables. Although, the exact sharpness parameter $\eta$ is unknown to us, we can put lower and upper bounds on the value of $\eta$ from the the observed values of witnesses $(W_{\text{AB}},W_{\text{AC}})$:
\begin{equation}
\eta \geq\sqrt{2}\left(2W_\text{AB}-1\right)\equiv \eta_\text{min},\qquad \eta\leq 2\sqrt{\left(2+\sqrt{2}-4W_\text{AC}\right)\left(2W_\text{AC}-1\right)}\equiv \eta_\text{max}
\end{equation}
We denote this region of compatible values of $\eta$ as $I(W_\text{AB},W_\text{AC})$. Since we consider extremal instruments the corresponding Kraus operators of Bob's operations are of the following form 
\begin{eqnarray}
&& K_{0|y}=\sqrt{\frac{1+\eta}{2}}\left(\frac{\openone}{2}+\frac{\hat{n}_y\cdot\vec{\sigma}}{2}\right)+\sqrt{\frac{1-\eta}{2}}\left(\frac{\openone}{2}-\frac{\hat{n}_y\cdot\vec{\sigma}}{2}\right),\\
&& K_{1|y}=\sqrt{\frac{1-\eta}{2}}\left(\frac{\openone}{2}+\frac{\hat{n}_y\cdot\vec{\sigma}}{2}\right)+\sqrt{\frac{1+\eta}{2}}\left(\frac{\openone}{2}-\frac{\hat{n}_y\cdot\vec{\sigma}}{2}\right).\nonumber
\end{eqnarray}
The average post-measurement state that Charlie receives from Bob can be found as follows
\begin{equation}
\bar{\rho}_x = \frac{1}{2}\sum_{y,b}\rho^{y,b}_x =  \frac{1}{2}\sum_{y,b}K_{b|y}\rho_x K_{b|y}^\dagger.
\end{equation}
We are interested in estimating the maximal possible purity of $\bar{\rho}_x$ that is compatible with the sharpness $\eta$, and observed witnesses $(W_{\text{AB}},W_{\text{AC}})$. To this aim we also optimize with respect to $\rho_x$, and hence we can omit the index $x$ and take it to be $\rho_x = \frac{\openone}{2}+\frac{\vec{a}\cdot\vec{\sigma}}{2}$. Performing simple calculations, one obtains the following explicit form of $\bar{\rho}$
\begin{equation}
\label{eq:post_m_state}
\bar{\rho} = \frac{\openone}{2}+\frac{1}{2}\sum_y\frac{1}{2}\Big(\sqrt{1-\eta^2}\vec{a}+(1-\sqrt{1-\eta^2})(\vec{a}\cdot\vec{n}_y)\vec{n}_y\Big)\cdot\vec{\sigma}. 
\end{equation}
To be more precise, we are interested not in the purity of  $\bar{\rho}$, but in the length of its Bloch vector $\vec{m}$, where $\bar{\rho} = \frac{\openone}{2}+\frac{\vec{m}\cdot\vec{\sigma}}{2}$. We can compute $|\vec{m}|$ as the norm of the vector that is multiplied by $\vec{\sigma}$ in Eq.~(\ref{eq:post_m_state}). These calculations result in the following expression
\begin{equation}
\label{eq:post_m_state_2}
|\vec{m}|^2 = 1-\eta^2+\left(\frac{3}{4}\eta^2+\frac{\sqrt{1-\eta^2}-1}{2}\right)\sum_y(\vec{a}\cdot\hat{n}_y)^2+\frac{(1-\sqrt{1-\eta^2})^2}{2}(\vec{a}\cdot\hat{n}_0)(\vec{a}\cdot\hat{n}_1)(\hat{n}_0\cdot\hat{n}_1).
\end{equation}
Due to the symmetry of the above expression with respect to $(\vec{a}\cdot\hat{n}_0)$ and $(\vec{a}\cdot\hat{n}_1)$, it is clear that the maximal value is achieved for $\vec{a} = \frac{\hat{n}_0+\hat{n}_1}{|\hat{n}_0+\hat{n}_1|}$, i.e. $(\vec{a}\cdot\hat{n}_0) = \frac{1}{\sqrt{2}}\sqrt{1+(\hat{n}_0\cdot\hat{n}_1)}$. Substituting the optimal $\vec{a}$ into Eq.~(\ref{eq:post_m_state_2}) gives the following short expression
\begin{equation}
|\vec{m}| = \frac{1}{2}|1+\sqrt{1-\eta^2}+(1-\sqrt{1-\eta^2})(\hat{n}_0\cdot\hat{n}_1)|.
\end{equation}
Now, we would like to establish the maximal overlap $(\hat{n}_0\cdot\hat{n}_1)$ compatible with the value of the witness $W_{\text{AB}}$ (the maximal $W_{\text{AB}} = \frac{1}{2}+\frac{1}{2\sqrt{2}}$ corresponds to mutually unbiased measurements, i.e. $(\hat{n}_0\cdot\hat{n}_1) = 0$). We refer to the results of Ref.~\cite{tavakoliKaniewski} that show that for a given QRAC value $W$ the following holds 
\begin{equation}
W\leq \frac{1}{2}+\frac{1}{16}\left(\sqrt{8\eta^2+2\mu}+\sqrt{8\eta^2-2\mu}\right),
\end{equation}
where $\mu=\Tr\left(\{B_0,B_1\}\right)=4\eta^2(\hat{n}_0\cdot\hat{n}_1)$. From here we directly obtain that 
\begin{equation}
|\hat{n}_0\cdot\hat{n}_1|\leq \frac{4W_{\text{AB}}-2}{\eta}\sqrt{2-\left(\frac{4W_{\text{AB}}-2}{\eta}\right)^2} = 2\frac{\eta_{\text{min}}}{\eta}\sqrt{1-\left(\frac{\eta_{\text{min}}}{\eta}\right)^2} \equiv f_{W_{\text{AB}}},
\end{equation}
and hence $|\vec{m}|\leq \frac{1}{2}|1+\sqrt{1-\eta^2}+(1-\sqrt{1-\eta^2})f_{W_{\text{AB}}}|$. 

Finally, we return to Eq.~(\ref{Wit}), where now $\vec{r}_0,\vec{r}_1$ correspond to the post-measurement sates that Bob sends to Charlie. We can bound the norm of both of these vectors by $|\vec{m}|$, which leads to
\begin{equation}
W_{\text{AC}}\leq \frac{1}{2}+\frac{1}{8}\left[|\vec{r}_0||\vec{c}_0+\vec{c}_1|+|\vec{r}_1||\vec{c}_0-\vec{c}_1|\right]\leq \frac{1}{2}+\frac{|1+\sqrt{1-\eta^2}+(1-\sqrt{1-\eta^2})f_{W_{\text{AB}}}|}{16}\left[|\vec{c}_0+\vec{c}_1|+|\vec{c}_0-\vec{c}_1|\right],
\end{equation}
where $\vec{c}_0$ and $\vec{c}_1$ are Bloch vectors of Charlie's measurements.  Since the degree of incompatibility is given by $D=|\vec{c}_0+\vec{c}_1|+|\vec{c}_0-\vec{c}_1|-2$, we can re-arrange this equation to obtain our final result
\begin{equation}
D\geq \frac{16W_\text{AC}-8}{1+\sqrt{1-\eta^2}+f_{W_\text{AB}}\left(1-\sqrt{1-\eta^2}\right)}-2.
\end{equation}
In order to estimate the above bound for the degree of incompatibility of Charlie's measurements, we should take the minimal value over the region of $\eta\in I(W_\text{AB},W_\text{AC})$.

\end{document}